\documentclass[referee]{raa} 
\usepackage{lscape}           
\usepackage{natbib}

\bibstyle{model2-names.bst}
\usepackage{xcolor} 
\usepackage{graphicx,times}             
\usepackage{amssymb,amsmath}
\bibpunct{(}{)}{;}{a}{}{,}

\usepackage[a4paper=true,dvipdfm=true,pagebackref=true]{hyperref}
\hypersetup{colorlinks = true, linkcolor = green, anchorcolor = red, citecolor = blue, filecolor = red, pagecolor = red, urlcolor = red}

\begin{document}

   \title{Testing the empirical relationship between Forbush decreases and cosmic ray diurnal anisotropy 
}

   \volnopage{Vol.0 (20xx) No.0, 000--000}      
   \setcounter{page}{1}          

   \author{Jibrin Adejoh Alhassan
      \inst{1*}
   \and Ogbonnaya Okike
      \inst{2}
   \and{Augustine Ejikeme Chukwude}
      \inst{1}
   }

   \institute{Department of Physics and Astronomy, University of Nigeria, Nsukka 400001, Nigeria \\
        \and
             Department of Industrial Physics, Ebonyi State University, Abakaliki 840001, Nigeria\\
{\it *Corr. Author:jibrin.alhassan@unn.edu.ng}\\ 
\vs\no
   {\small Received~~20xx month day; accepted~~20xx~~month day}}

\abstract{ The abrupt aperiodic modulation of  cosmic ray (CR) flux intensity, often referred to as Forbush decrease (FD), plays a significant role in our understanding of the Sun-Earth electrodynamics. Accurate and precise determination of FD magnitude and timing are among the intractable problems in FD-based analysis. FD identification is complicated by CR diurnal anisotropy. CR anisotropy can  increase or reduce the number and amplitude of FDs. It is therefore important to remove its  contributions  from CR raw data before FD identification. Recently, an attempt was made, using  a combination of Fourier transformed technique and FD-location machine  to address this. Thus, two FD catalogs and amplitude diurnal variation (ADV) were calculated from filtered (FD1 and ADV) and raw (FD2) CR data. In the current work, we test the  empirical  relationship between FD1, FD2, ADV  and solar-geophysical  characteristics.  Our analysis shows that two types of magnetic fields-interplanetary (IMF) and geomagnetic (Dst) govern the  evolution of CR flux intensity reductions.  
\keywords{methods: data analysis - 
methods: statistical- Sun: coronal mass ejections (CMEs) - (Sun:) solar - terrestrial relations - (Sun:) solar wind - (ISM:) cosmic rays}
}

   \authorrunning{Alhassan, Okike \and Chukwude}  
   \titlerunning{Forbush Decreases and Cosmic Ray Anisotropy}   

   \maketitle

\section{Introduction}           
\label{sect:intro}
Galactic cosmic ray (GCR) intensity flux which is believed to be  modulated by solar wind  interplanetary magnetic field (IMF) structure, among several other agents, include  periodic and  aperiodic components.  The periodic category include cosmic ray (CR) diurnal anisotropy \citep{ogbos:21}, 27-d, and 11-year long term modulations \citep{oh:08}. CR diurnal anisotropy may be viewed as  a periodic, short-term variation in CR flux. It is the portion of the total CR intensity variation with 24 hr periodicity resulting from the Earth's rotation about its axis coupled with the changes of asymtotic cone of acceptance of neutron monitors (NMs) \citep{Lockwood:1971}. One of  the abrupt time-intensity changes of CR flux is Forbush decrease (FD) event which is named after the pioneer observer of the phenomenon, \citep{fo:38}. FD  is a non-periodic sudden reduction in CR intensity flux caused by interplanetary  disturbances (IPDs) in the form of  magnetic field enhancements in interplanetary space and high velocity solar wind \citep{fo:38, ba:75, rao:76}. There are two main kinds of IPDs-sporadic and recurrent. Sporadic IPDs are caused by coronal mass ejections (CMEs) and their IP version-interplanetary coronal mass ejections (ICMEs) while recurrent IPDs are associated with heliospheric current sheet  and high-speed plasma flows from coronal holes (CHs) which co-rotate with the Sun \citep{belov:01, be:2014, alh:2021a}. FDs from CHs are known to have small magnitudes while those caused by CMEs have  large magnitude signatures \citep{ Lockwood:1971, avbelov:2009, Melkumyan:2019}.

Large FDs from isolated  NMs appear to be relatively easier to detect than  small events. This is attributed to the elusive nature of weak events believed to be caused by the  masking tendencies of CR diurnal anisotropies \citep{ba:75}. Apart from the challenges  associated with weak event detection, \citet{ogbona:2020} observed that  the inherent CR  effects that range from  enhanced diurnal anisotropies, signal superposition, periodicities, cycles to short-term random variations on the amplitude and timing of FDs \citep{ca:96, cane:2003, oh:08, rich:2011} are scarcely removed from raw CR data. This is because the well-known manual method of FD detection is not capable in   handling  the superposed effects of the "unwanted signals"  on CR data \citep{ok:2021}.
If the contributions of CR diurnal anisotropies are not considered  before FD identification, some of the events selected might just  be enhanced diurnal CR anisotropy, pre-increases or pre-decreases that happen before the actual CR depression \citep{ok:2019}. Fully automated FD identification that clearly deal with these daunting issues have become the interest of recent works \citep[e.g.][]{ok:2019, ogbona:2020, ogbos:21}.

In the study of empirical implication of conducting Chree analysis with data from isolated NM stations, \citet{ok:2019} developed a FD-location program which is based on Fourier transformation. Raw CR data are first transformed into sinusoidal waves. The imaginary part that handles the daily and diurnal variations are discarded. The real part serves as the input signal to the FD-location  program. The FD-location program involves several different calculations. Some subroutines detect both small and large transient intensity reductions (minima/pits) as well as increases (maxima/peaks) in cosmic ray (CR) data. Other sub-modules calculate event magnitude, timing and cataloging of the events identified. The subroutines that track increases in CR flux in the form of solar energetic particles (SEPs) and ground level enhancements (GLEs) are disabled while only reductions in CR flux  are selected. While Fourier transformation can remove the slow-moving signal in any data,  a step beyond Fourier decomposition that can  calculate the FD event date and magnitude is  demonstrated in the referred publication. 

 A direct application of Fourier transform techniques to handle the enhanced diurnal CR wave trains that accompany FDs is a subject of research interest. \citet{ogbona:2020} carried out a detailed study of simultaneous and non-simultaneous FDs with focus on the implications of CR diurnal oscillations on FDs at different geographical locations at Earth. This publication developed a FD location  algorithm that was used  to select FDs from both raw (unprocessed) and Fourier transformed CR data. The code which uses static mean accepts raw CR data as input signal and is able to calculate both the event time and amplitude  concurrently. In addition to the R program, the paper employed Fast Fourier transformation (FFT) in order  to decompose the signals into their respective frequency domains to account for the CR diurnal anisotropy that occur at the time of FDs. For the first time, the algorithm selected two FD catalogs-FD1 (FD from preliminary processed data) and FD2 (FD from unprocessed data). The result of their analysis shows that the amplitude of CR diurnal wave is about 13 or 20 times the magnitude  predicted by \citet{ax:1965, mc:1965} but consistent with   the high amplitude ($\approx$ 10\%)  by \citet{be:08}.  

\begin{center}
	\begin{figure}[hb]
	\centering
	\includegraphics[width=10cm]{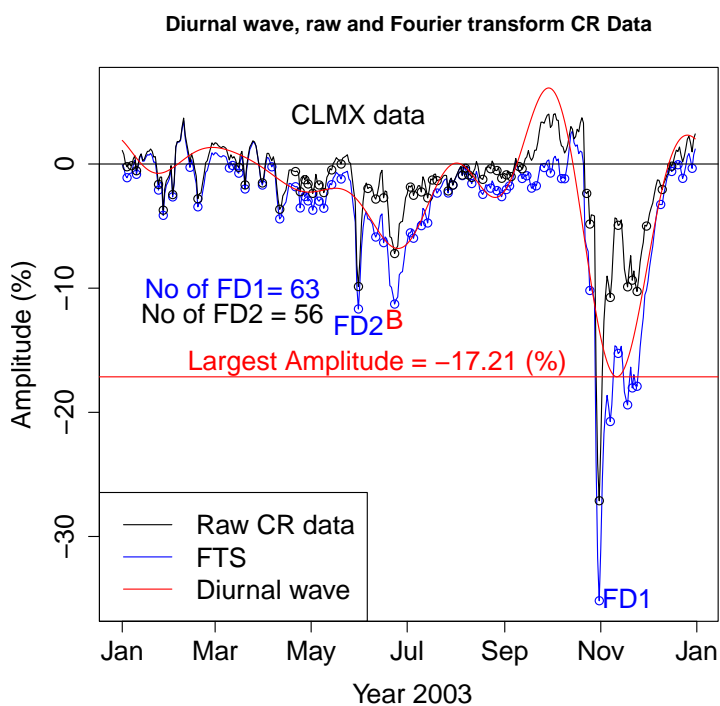}
	\caption{Comparison of the amplitude of CR diurnal anisotropy with the raw CR data, the Fourier transformed signal, and the associated FDs for the year 2003 at CLMX station (adapted from \citep{ogbos:21}).}
	\label{Figure 1}
	 \end{figure}
	\end{center}

With some technical improvements in FD location algorithm, \citet{ogbos:21}  adjusted for the influence of  anisotropy in CR data as well as removal of the solar cycle variations from the observed  amplitude of FDs  at Climax NM station. This algorithm allows for accurate calibration and ranking of FDs. A comparison of the amplitude of CR diurnal anisotropy with the raw CR data, the Fourier transformed signal, and the associated FDs for the year 2003 at CLMX station was demonstrated (see Figure \ref{Figure 1}). The low velocity and high velocity signals were separated from the raw CR data using the FFT technique. The empirical connection between CR diurnal oscillation and FDs detected from unprocessed (FD2) and Fourier transformed (FD1) CR data are determined utilizing CR data from Climax NM station for the period 1953-2006. \citet{ogbos:21} found strong and statistically  significant correlations between FD1, FD2  and ADV. The correlation coefficient between FD1 and ADV tend to be higher than those of FD2 and ADV. This underscores the proposition that CR anisotropy is an integral part of CR depressions. 

Using numerical filtering techniques on CR data from two isolated NM stations Apatity (APTY) and Mt.Washington (MTWS) during high solar activity in 1972, \citet{okj:21}, hereafter, Paper I, demonstrated that the low frequency component of CR flux in which CR anisotropy is coded could be disentangled from the rapidly varying portion that contains the FDs. The high velocity signal component is then passed on to an FD location software for accurate event timing and amplitude calculation (FD1)  while the magnitudes of the CR diurnal anisotropies (ADV)  are obtained from the low-energy spectrum of the raw CR data. An FD location algorithm was further used to estimate the amplitude of FDs from unprocessed CR data (FD2). Thus, two FD catalogs and amplitude diurnal variation (ADV) were calculated from Fourier decomposed (FD1 and ADV) and raw (FD2) CR data. A correlation coefficient of $\approx 0.98$ was found between FDs at APTY and MTWS for both the raw and the transformed data. The high correlation between the FD amplitudes at the two stations may be an indication of the efficiency of the algorithms deployed in Paper I.
Presently, there is no FD selection approach that can adequately solve all the problems associated with FD detection, hence the  need to validate any selected FD  list \citep{ok:2020}. Obtaining valid FDs is very crucial since it is suggested to be an important tool used to  examine the electrodynamics of the solar-terrestrial connection (see Paper I). 

The two general approaches to  FD identification from CR data include detection of FDs without recourse to solar-wind characteristics \citep[e.g.][]{pu:95-2, har:2010} and detection of FDs from  CR data with solar-wind parameters \citep[e.g.][]{avbelov:2009,rami:2013}. The association between FD and speculated causative agents like CME, ICME, solar wind speed (SWS), geomagnetic storm index (Dst), interplanetary magnetic field (IMF) etc is not yet well understood. In the present work, the link between  FD1, FD2, ADV catalogs from APTY and MTWS NM stations and the associated solar-wind data is tested to determine whether the events are real or spurious. 
 
\section{Data}
The two FD catalogs: FD1-from Fourier transformed and FD2- from raw CR data with the corresponding daily  ADV based on  APTY (Longitude = 43.28$^{\circ}$N, Latitude = 42.69$^{\circ}$E, $R_{c}$ = 5.6 GV and Altitude = 1700m ) and MTWS (Longitude = 44.27$^{\circ}$N, Latitude = -71.30$^{\circ}$W, $R_{c}$ = 1.46 GV and Altitude = 1909m) high  neutron monitor detectors are taken from Paper I. IMF, SWS and  Dst data are downloaded from omniweb.gsfc.nasa.gov/html/ow data.html.
\section{Results and Discussions}

\subsection{FD1, FD2 and ADV against Solar-geophysical parameters}
\begin{figure}
  \resizebox{\hsize}{!}{\includegraphics{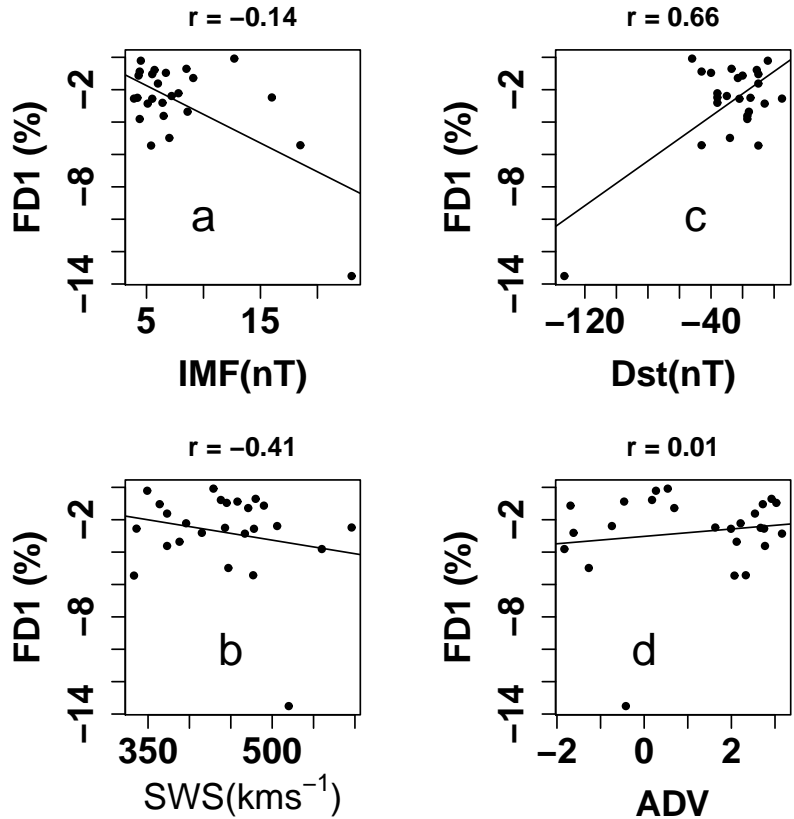}}
  \caption{Scatter Plots of Magnitude of $FD1_{APTY}$  versus Solar Wind Parameters and  CR Diurnal Anisotropy} 
  \label{Figure 2}
\end{figure}

\begin{figure}
  \resizebox{\hsize}{!}{\includegraphics{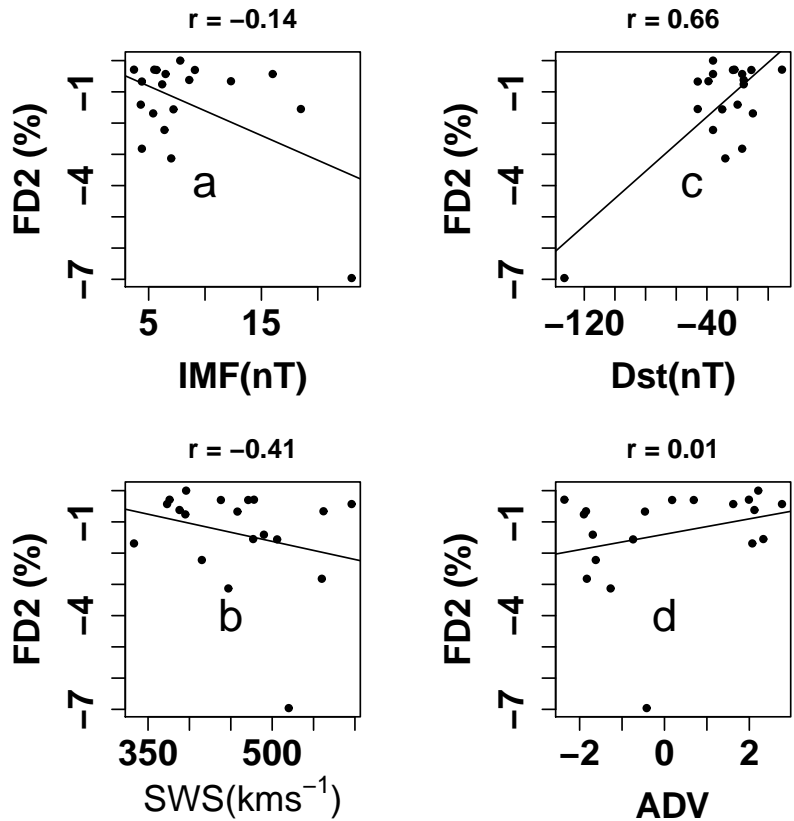}}
  \caption{Graph of Amplitude of $FD2_{APTY}$ against Solar Wind Parameters and  CR Diurnal Anisotropy}   
  \label{Figure 3}
\end{figure}

\begin{figure}
  \resizebox{\hsize}{!}{\includegraphics{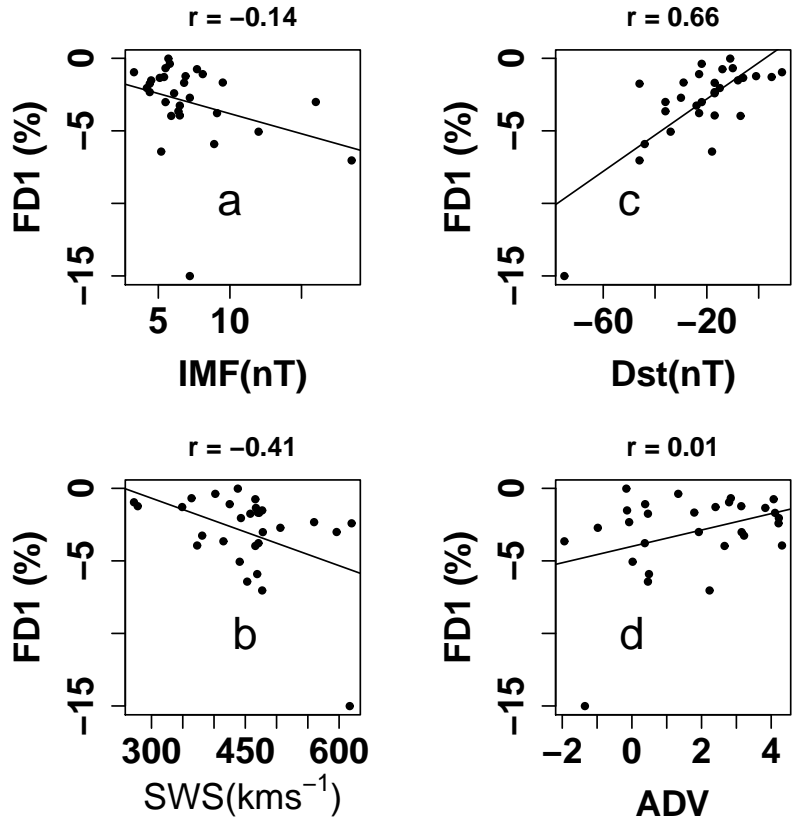}}
  \caption{Plots of  $FD1_{MTWS}$ Magnitude, Solar Wind Parameters and Magnitude of CR Diurnal wave } 
  \label{Figure 4}
\end{figure}

\begin{figure}
  \resizebox{\hsize}{!}{\includegraphics{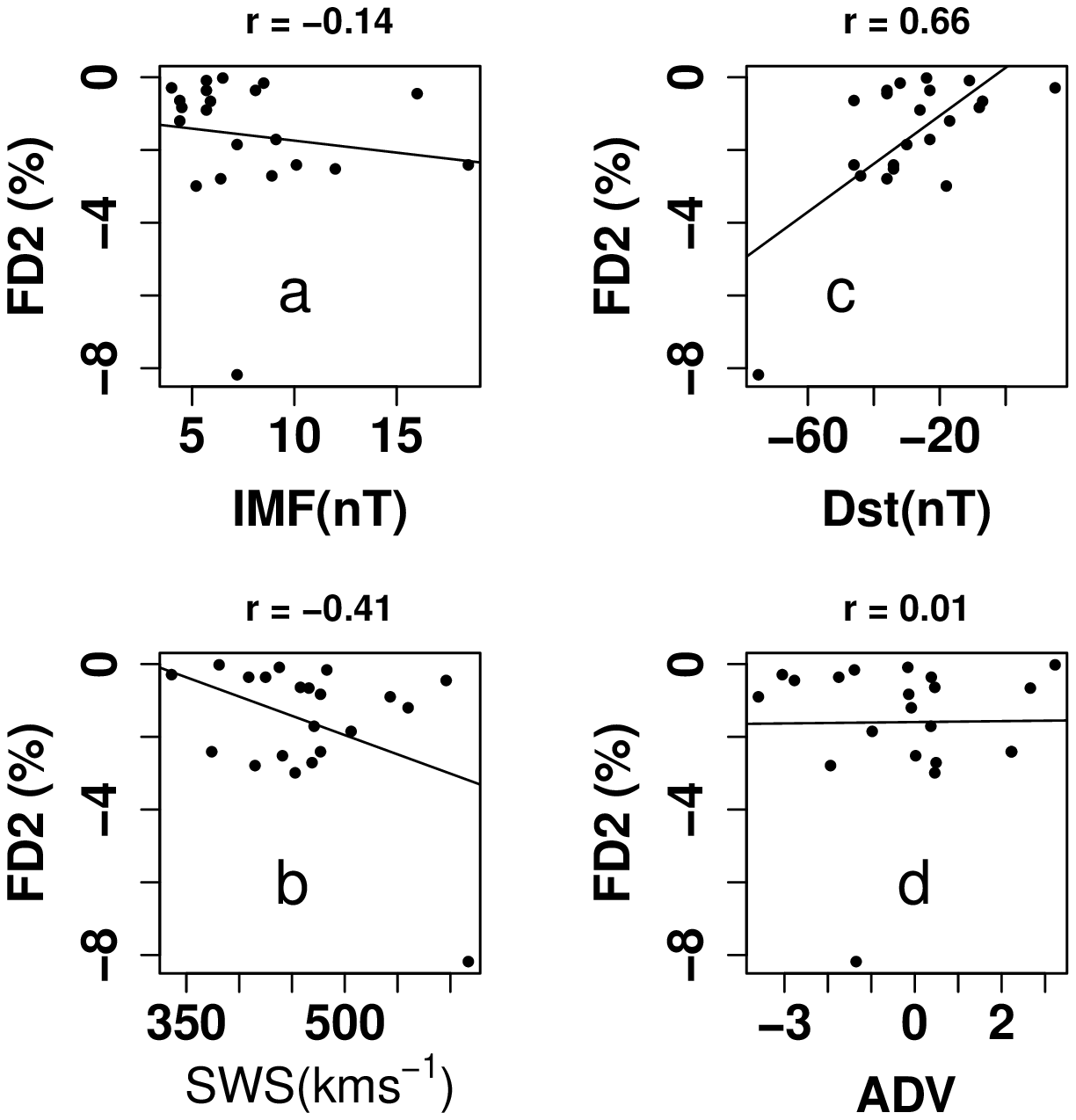}}
  \caption{Correlation between  $FD2_{MTWS}$ Event sizes, Solar Wind Parameters and Magnitude of CR Diurnal Wave}  
  \label{Figure 5}
\end{figure}

The  magnitude of FDs and ADV  for APTY   and   MTWS   calculated by the FD location algorithm taken from  table 2 of  Paper 1 with  the corresponding solar wind parameters are as presented respectively  in  Table \ref{table 1}, Table \ref{table 2}, Table \ref{table 3} and Table \ref{table 4}. 
The regression and correlation results of the two FD data sets, ADV and solar wind data for APTY and MTWS detectors are given in  Table \ref{table 5}, Table \ref{table 6}, Table \ref{table 7} and Table \ref{table 8}. The statistical significance test of the correlation coefficient (r) is based on the $F_{statistic}$ indicated on the regression result Tables. $F_{statistic}$ refers to ratio of two variances that test significance of regression. The plots of the APTY and MTWS FD1 and FD2  against IMF, SWS, Dst and ADV are displayed, respectively, in Figures \ref{Figure 2}, \ref{Figure 3}, \ref{Figure 4} and \ref{Figure 5}. The corresponding regression equations \ref{eqn1}, \ref{eqn2}, and \ref{eqn3} (all not shown) for the graphs reflect the correlation results.

The coefficient of determination ($R^2$), r and  chance probability (p-values) for FD$-$IMF diagrams of Figures \ref{Figure 2}a, \ref{Figure 3}a, \ref{Figure 4}a and \ref{Figure 5}a are given in Tables \ref{table 5}, \ref{table 6}, \ref{table 7} and \ref{table 8} respectively. Whereas the FD1$-$IMF and FD2$-$IMF respective correlation coefficients of -0.63 and -0.51 at APTY are statistically significant respectively at 99\% and 95\% confidence, the correlation results for MTWS of -0.32 and -0.14 are statistically non-significant. 
The regression analysis of FD$-$IMF relation at APTY and MTWS may imply that the FD from transformed and raw CR data are dependent on IMF intensity. 

The FD-SWS plots displayed in Figures \ref{Figure 2}b, \ref{Figure 3}b, \ref{Figure 4}b and \ref{Figure 5}b, yield  correlation coefficients of -0.02, -0.26; -0.44, -0.41 respectively for FD1$-$SWS and FD2$-$SWS connections at the two stations. 
 
The FD-Dst relation is very striking. Compared to other parameters, the correlation coefficients for the FD-Dst relation are statistically significant at both stations for all the FD data sets. It is shown to be  at 99\% and 95\% confidence levels. This suggests  that the CR variations at the two stations are  driven by geomagnetic storm time  activity. Quantitatively, from the $R^2$ values,  Dst index appear to account for more than half of  CR depressions in the present data. The plots of FD-ADV relation seem not  to show any statistically significant correlations for the two data sets at the two stations. 

\begin{equation} \label{eqn1} FD1_{APTY} = 4.79\pm 1.10+(1.11\pm 0.29)IMF
\end{equation}
\begin{equation} \label{eqn2} FD1_{APTY} = 426.85\pm 19.49+(5.21\pm 0.51)SWS
\end{equation}
\begin{equation} \label{eqn3}  FD1_{APTY} = -9.26\pm 5.76+(6.57\pm 1.50)Dst
\end{equation}

We have studied  the association between two separate FD catalogs (FD1 and FD2), solar-geophysical parameters and the associated magnitudes of diurnal anisotropies observed at APTY and MTWS stations during the year 1972 which is a period of high solar activity. The results of FD1$-$IMF, FD1$-$Dst  at APTY; FD1$-$SWS, FD1$-$Dst at MTWS which are statistically significant are consistent with the submission of  \citet{ok:2019a} that found statistically significant correlations for FD-IMF (r = -0.39), FD$-$SWS (r = -0.71) and FD$-$Dst (r = 0.45) for the processed CR data.  FD1$-$SWS at APTY and FD1$-$IMF at MTWS that are statistically non-significant, 
do not reflect their findings. 

Using  a total of 17 and 68 FD events respectively, \citet{ka:10} and \citet{lin:2016} investigated  the connection between FD amplitude and Dst index but did not  find any discernible pattern between the two parameters. \citet{belov:01} found a correlation coefficient  r $<$ 0.42 between FD and Dst. We find statistically significant FD2$-$Dst correlations contrary to their reports. This could be an indication of  the differences in the semi-automated and the present fully automated FD event identification approaches \citep{alh:2021a}. \citet{ok:2020},  from a critique of the traditional manual technique of determination of the magnitude of FDs, reported  correlation coefficients for FD$-$Dst and FD$-$SWS relations at three CR stations: ESOI station (ESOI), McMurdo (MCMD) and Thule (THUL)  respectively as  0.18, 0.34, 0.32 and  0.00, -0.11, -0.12. The non-statistically significant results we obtained  here for FD2$-$SWS of r = -0.26 for APTY and -0.41 for MTWS  is at variance  with their result. 
Our result suggests that different mechanisms might be responsible for the FDs and SWS. The FD2$-$Dst results of  r = 0.76 and r = 0.66 respectively for APTY and MTWS  NMs reported in the current analysis agrees with their finding.
 
In an important review of FDs, \citet{Lockwood:1971} suggested that IMF and Dst are responsible for the high-frequency modulation of CRs. The  strong connection between the amplitude of FD and Dst activity for the two data sets at 99\% and 95\%  confidence level and FD$-$IMF relation at APTY for the two FD catalogs significant at 99\% and 95\% confidence level reported here is consistent with their proposition. In the context of previous publications, we report that these two types of magnetic fields (the interplanetary (IMF) and geomagnetic (Dst)) could be  the causative agents of the high frequency variation of CRs. Recently, \citet{alh:2021b, alh:2021a} found statistically significant correlations between FDs from raw CR data, IMF, SWS and Dst. The current results for FD2$-$IMF, FD2$-$Dst for APTY and FD2$-$Dst for MTWS reflect the finding of these authors. Our results of FD2$-$SWS at APTY and FD2$-$IMF, FD2$-$SWS at MTWS is contrary to their submissions. This present findings show that the mechanisms responsible for FD, IMF and SWS are not the same and also that IMF and SWS  may not play significant roles in CR modulations when unprocessed data are considered as previously reported. This may also be due to the masking effect of diurnal anisotropy on the CR data. 
  
CR anisotropy has been identified as an important signal in CR flux intensity reductions by \citet{ogbos:21}.  For this reason, a linear relationship between the amplitude of FDs and the magnitude  of the CR diurnal wave should be envisaged. We examined this relationship and found  that there exist no statistically significant correlation between FD1, FD2 and the amplitude of the diurnal oscillation. Our regression analysis  does not reflect the results of \citet{ogbos:21} in which significant correlation were reported especially between FD1 and ADV. This trend is understandable in the light of the findings of Paper I. Paper I demonstrated that it is difficult to determine a pattern between FDs  and ADV. This is due to the fact that in some cases, anisotropy tend to reduce the magnitude of FDs or enhance it. In some other cases, the effect seems negligible. The association between them is rather very complex and does not seem to have definite pattern. 
\subsection{GCR modulation dependence on rigidity/NM efficiency} 
It has  been reported  that asymptotic cone of acceptance and geomagnetic cutoff rigidity determines whether a given neutron monitor observes an increasing or decreasing GCR flux \citep{sm:03-2, an:09-2}. This has been attributed to the fact that the distribution of GCR flux over the Earth is asymetrical, but could be the result of the association between the IMF and the geomagnetic field at a particular location on Earth. We test the proposition that the monitors with the lowest vertical cutoff rigidity could be more sensitive to variation in counting rate with  FDs observed at APTY and MTWS with different rigidities taken from table 2 of Paper I. For the largest event of 05 August 1972, the magnitude at APTY and MTWS are -25.49\% and -29.22\% respectively. The event of 18 June 1972, -7.50\% and -8.57\% magnitudes were calculated for APTY and MTWS respectively. For the least event on 02 Jan 1972, the amplitude at APTY is -0.78\% while that at MTWS is -0.02\%. Examining the corresponding events at the two stations, we observe that on average, the NM at MTWS that is characterized by low vertical cutoff rigidity is more sensitive  to CR intensity variations during FDs than the detector at APTY with higher rigidity. This result is consistent with the findings of \citet{ok:2020}.

\section{Conclusions}
Raw CR data are characterized by high variability and different  superposed signals of dissimilar periodicities, cycles and recurrences  such as  FDs, diurnal anisotropies, solar energetic particles (SEPs) and GLEs \citep{ogbos:21}. The measurement of the magnitude of FDs and the accurate timing  of its occurrence will be difficult to achieve  with the manual FD detection method.  Application of Fourier transform to isolated NM data is capable of filtering out undesirable signals superposed on raw CR counts. This  led to the identification of two FD catalogs and amplitude diurnal variation (ADV)  from filtered (FD1 and ADV) and raw (FD2) CR data by Paper I. Establishing the link between FD and solar-geophysical activity indices as a means of validating FD list is still poorly understood as existing publications yield conflicting results. The conflicting submissions may be argued to stem from different FD data identified by investigators using different NM data. 

The two dimensional regression analysis carried out in this study reveal that 
two types of magnetic fields-interplanetary (IMF) and geomagnetic (Dst) appear to be responsible for FD detection as evident from FD$-$IMF/Dst  statistically significant correlations. We did not find evidence of significant FD$-$ADV correlation from the two FD catalogs. This could be due to the  complex link between FD and ADV. 
\onecolumn
\begin{table}[ht]
\caption{ \textbf{$APTY_{FD1}$, ADV1 and Solar Wind data. "Date"  for time of maximal CR decreases, "FD1" represents magnitude of FD from transformed APTY CR data and "ADV1"  for magnitude of diurnal anisotropy corresponding to FD1}}
\label{table 1}
\centering
\begin{tabular}{rlrrrrr}
  \hline
 Order & Date & IMF & SWS & Dst & FD1(\%) & ADV1(\%) \\ 
  \hline
1 & 1972-01-02 & 5.70 & 438 & -11 & -0.78 & 0.18 \\ 
  2 & 1972-01-30 & 4.30 & 490 & -20 & -1.13 & -1.69 \\ 
  3 & 1972-02-15 & 4.40 & 560 & -17 & -3.81 & -1.83 \\ 
  4 & 1972-02-20 & 7.00 & 447 & -28 & -4.98 & -1.27 \\ 
  5 & 1972-02-25 & 4.40 & 458 & -46 & -0.88 & -0.46 \\ 
  6 & 1972-03-07 & 16.00 & 596 & -36 & -2.48 & 1.63 \\ 
  7 & 1972-03-10 & 8.60 & 388 & -16 & -3.36 & 2.12 \\ 
  8 & 1972-03-15 & 3.90 & 336 &   5 & -2.55 & 2.75 \\ 
  9 & 1972-03-17 & 8.50 & 480 & -27 & -0.71 & 2.92 \\ 
  10 & 1972-03-30 & 6.70 & 445 & -40 & -0.96 & 3.03 \\ 
  11 & 1972-04-07 & 4.20 & 443 & -15 & -2.50 & 2.67 \\ 
  12 & 1972-04-12 & 6.00 & 373 & -10 & -1.62 & 2.54 \\ 
  13 & 1972-04-23 & 6.50 & 373 & -17 & -3.62 & 2.77 \\ 
  14 & 1972-05-06 & 5.10 & 467 &  -6 & -2.86 & 3.16 \\ 
  15 & 1972-05-13 & 5.50 & 364 & -10 & -1.04 & 2.72 \\ 
  16 & 1972-05-16 & 18.50 & 477 & -46 & -5.43 & 2.33 \\ 
  17 & 1972-06-23 & 6.40 & 415 & -36 & -2.81 & -1.62 \\ 
  18 & 1972-06-28 & 7.20 & 506 & -30 & -2.39 & -0.74 \\ 
  19 & 1972-07-04 & 4.50 & 349 &  -4 & -0.21 & 0.27 \\ 
  20 & 1972-10-11 & 7.80 & 396 & -36 & -2.22 & 2.21 \\ 
  21 & 1972-10-19 & 12.70 & 429 & -52 & -0.08 & 0.54 \\ 
  22 & 1972-11-01 & 23.00 & 520 & -133 & -13.51 & -0.42 \\ 
  23 & 1972-11-26 & 5.50 & 478 & -22 & -2.56 & 1.99 \\ 
  24 & 1972-11-30 & 5.40 & 333 & -10 & -5.45 & 2.07 \\ 
  25 & 1972-12-23 & 9.10 & 471 & -23 & -1.28 & 0.69 \\ 
   \hline
\end{tabular}
\end{table}
 
\begin{table}[ht]
\caption{ \textbf{$APTY_{FD2}$, ADV2 and Solar Wind data. "Date"  for time of maximal CR decreases, "FD2" represents magnitude of FD from raw APTY CR data and "ADV2"  for magnitude of diurnal anisotropy corresponding to FD2}}
\label{table 2}
\centering
\begin{tabular}{rlrrrrr}
  \hline
 Order & Date & IMF & SWS & Dst & FD2(\%) & ADV2(\%)  \\ 
  \hline
1 & 1972-01-02 & 5.70 & 438 & -11 & -0.30 & 0.18 \\ 
  2 & 1972-01-30 & 4.30 & 490 & -20 & -1.41 & -1.69 \\ 
  3 & 1972-02-02 & 6.20 & 395 & -16 & -0.76 & -1.90 \\ 
  4 & 1972-02-15 & 4.40 & 560 & -17 & -2.82 & -1.83 \\ 
  5 & 1972-02-20 & 7.00 & 447 & -28 & -3.13 & -1.27 \\ 
  6 & 1972-02-25 & 4.40 & 458 & -46 & -0.67 & -0.46 \\ 
  7 & 1972-03-07 & 16.00 & 596 & -36 & -0.43 & 1.62 \\ 
  8 & 1972-03-10 & 8.60 & 388 & -16 & -0.62 & 2.12 \\ 
  9 & 1972-04-23 & 6.50 & 373 & -17 & -0.43 & 2.77 \\ 
  10 & 1972-05-16 & 18.50 & 477 & -46 & -1.55 & 2.33 \\ 
  11 & 1972-06-10 & 3.70 & 376 &   9 & -0.29 & -2.36 \\ 
  12 & 1972-06-23 & 6.40 & 415 & -36 & -2.22 & -1.62 \\ 
  13 & 1972-06-28 & 7.20 & 506 & -30 & -1.56 & -0.74 \\ 
  14 & 1972-07-25 & 12.30 & 562 & -39 & -0.66 & -1.85 \\ 
  15 & 1972-10-11 & 7.80 & 396 & -36 & 0.00 & 2.21 \\ 
  16 & 1972-11-01 & 23.00 & 520 & -133 & -6.96 & -0.42 \\ 
  17 & 1972-11-26 & 5.50 & 478 & -22 & -0.29 & 1.99 \\ 
  18 & 1972-11-30 & 5.40 & 333 & -10 & -1.69 & 2.07 \\ 
  19 & 1972-12-23 & 9.10 & 471 & -23 & -0.30 & 0.69 \\ 
   \hline
\end{tabular}
\end{table}
 
\begin{table}[ht]
\caption{ \textbf{$MTWS_{FD1}$, ADV1 and Solar Wind data. "Date"  for time of maximal CR decreases, "FD1" represents magnitude of FD from transformed MTWS CR data and "ADV1"  for magnitude of diurnal anisotropy associated with FD1}}
\label{table 3}
\centering
\begin{tabular}{rlrrrrr}
  \hline
 Order & Date & IMF & SWS & Dst & FD1(\%) & ADV1(\%)  \\ 
  \hline
1 & 1972-01-02 & 5.70 & 438 & -11 & -0.02 & -0.16 \\ 
  2 & 1972-01-04 & 4.50 & 477 &  -8 & -1.52 & -0.14 \\ 
  3 & 1972-01-20 & 5.20 & 453 & -18 & -6.43 & 0.46 \\ 
  4 & 1972-01-22 & 8.90 & 469 & -44 & -5.91 & 0.49 \\ 
  5 & 1972-02-15 & 4.40 & 560 & -17 & -2.33 & -0.08 \\ 
  6 & 1972-02-19 & 12.00 & 441 & -34 & -5.06 & 0.02 \\ 
  7 & 1972-02-25 & 4.40 & 458 & -46 & -1.75 & 0.46 \\ 
  8 & 1972-03-03 & 5.80 & 402 & -22 & -0.38 & 1.33 \\ 
  9 & 1972-03-07 & 16.00 & 596 & -36 & -3.01 & 1.92 \\ 
  10 & 1972-03-12 & 5.90 & 466 &  -7 & -3.97 & 2.66 \\ 
  11 & 1972-03-29 & 6.80 & 472 & -17 & -1.69 & 4.11 \\ 
  12 & 1972-04-05 & 6.10 & 620 & -17 & -2.41 & 4.21 \\ 
  13 & 1972-04-07 & 4.20 & 443 & -15 & -2.05 & 4.22 \\ 
  14 & 1972-04-23 & 6.50 & 373 & -17 & -3.94 & 4.31 \\ 
  15 & 1972-05-03 & 7.70 & 466 & -14 & -0.75 & 4.07 \\ 
  16 & 1972-05-06 & 5.10 & 467 &  -6 & -1.35 & 3.83 \\ 
  17 & 1972-05-13 & 5.50 & 364 & -10 & -0.68 & 2.84 \\ 
  18 & 1972-05-16 & 18.50 & 477 & -46 & -7.04 & 2.23 \\ 
  19 & 1972-06-23 & 6.40 & 415 & -36 & -3.65 & -1.94 \\ 
  20 & 1972-06-28 & 7.20 & 506 & -30 & -2.72 & -0.98 \\ 
  21 & 1972-10-12 & 8.10 & 425 & -23 & -1.09 & 0.38 \\ 
  22 & 1972-11-02 & 7.20 & 617 & -75 & -15.01 & -1.35 \\ 
  23 & 1972-11-26 & 5.50 & 478 & -22 & -3.02 & 3.15 \\ 
  24 & 1972-11-29 & 6.50 & 381 & -24 & -3.26 & 3.22 \\ 
  25 & 1972-12-02 & 6.90 & 278 &  -1 & -1.23 & 3.14 \\ 
  26 & 1972-12-06 & 3.30 & 272 &   9 & -0.96 & 2.79 \\ 
  27 & 1972-12-09 & 5.40 & 349 &   5 & -1.29 & 2.40 \\ 
  28 & 1972-12-13 & 9.50 & 470 & -29 & -1.67 & 1.79 \\ 
  29 & 1972-12-23 & 9.10 & 471 & -23 & -3.78 & 0.37 \\ 
   \hline
\end{tabular}
\end{table}
 
\begin{table}[ht]
\caption{ \textbf{$MTWS_{FD2}$, ADV2 and Solar Wind data. "Date"  for time of maximal CR decreases, "FD2" represents magnitude of FD from raw MTWS CR data and "ADV2"  for magnitude of diurnal anisotropy associated with FD2}}
\label{table 4}
\centering
\begin{tabular}{rlrrrrr}
  \hline
 Order & Date & IMF & SWS & Dst & FD2(\%) & ADV2(\%)  \\ 
  \hline
1 & 1972-01-02 & 5.70 & 438 & -11 & -0.09 & -0.16 \\ 
  2 & 1972-01-04 & 4.50 & 477 &  -8 & -0.83 & -0.14 \\ 
  3 & 1972-01-20 & 5.20 & 453 & -18 & -2.99 & 0.46 \\ 
  4 & 1972-01-22 & 8.90 & 469 & -44 & -2.71 & 0.49 \\ 
  5 & 1972-02-15 & 4.40 & 560 & -17 & -1.20 & -0.08 \\ 
  6 & 1972-02-19 & 12.00 & 441 & -34 & -2.52 & 0.02 \\ 
  7 & 1972-02-25 & 4.40 & 458 & -46 & -0.64 & 0.46 \\ 
  8 & 1972-03-07 & 16.00 & 596 & -36 & -0.45 & -2.77 \\ 
  9 & 1972-03-12 & 5.90 & 466 &  -7 & -0.66 & 2.66 \\ 
  10 & 1972-03-16 & 10.10 & 374 & -34 & -2.41 & 2.23 \\ 
  11 & 1972-05-16 & 18.50 & 477 & -46 & -2.41 & 2.22 \\ 
  12 & 1972-06-11 & 4.00 & 336 &  15 & -0.29 & -3.05 \\ 
  13 & 1972-06-23 & 6.40 & 415 & -36 & -2.79 & -1.94 \\ 
  14 & 1972-06-28 & 7.20 & 506 & -30 & -1.85 & -0.98 \\ 
  15 & 1972-07-26 & 5.70 & 543 & -26 & -0.90 & -3.60 \\ 
  16 & 1972-10-12 & 8.10 & 425 & -23 & -0.36 & 0.38 \\ 
  17 & 1972-10-20 & 8.50 & 483 & -32 & -0.16 & -1.39 \\ 
  18 & 1972-10-23 & 5.70 & 409 & -36 & -0.36 & -1.75 \\ 
  19 & 1972-11-02 & 7.20 & 617 & -75 & -8.18 & -1.35 \\ 
  20 & 1972-11-29 & 6.50 & 381 & -24 & -0.02 & 3.23 \\ 
  21 & 1972-12-23 & 9.10 & 471 & -23 & -1.71 & 0.37 \\ 
   \hline
\end{tabular}
\end{table}

\begin{table}[ht]
\caption{ \textbf{ APTY FD1 Regression Results with other Parameters; $F_{critical}$ = 7.88 (99\% confidence). None statistically significant correlations are marked with *. "Order", stands for the sequence of parameters, $R{^2}$ denotes coefficient of determination, r represents coefficient of correlation, $p-values$ is the chance probability, $F_{statistic}$ refers to ratio of two variances that test significance of regression}}. 
\label{table 5}
\centering
\begin{tabular}{rlrrrr}
  \hline
Order & Parameters & $R{^2}$ & r & $p-values$ & $F_{statistic}$ \\ 
  \hline
1 & FD1-IMF & 0.40 & -0.63 & $7.62\times 10^{-04}$ & 15.04 \\ 
  2 & FD1-SWS & 0.04 & -0.02* & 0.32 & 1.05 \\ 
  3 & FD1-Dst & 0.45 & 0.67 & $2.24\times 10^{-04}$ & 19.09 \\ 
  4 & FD1-ADV & 0.02 & 0.14* & 0.49 & 0.48 \\ 
   \hline
\end{tabular}
\end{table}

\begin{table}[ht]
\caption{ \textbf{ APTY FD2 Regression Results with other Variables; $F_{critical}$ = 4.45; (95\% confidence). None statistically significant correlations are marked with *. Parameters are as defined in Table \ref{table 5}}}
\label{table 6}
\centering
\begin{tabular}{rlrrrr}
  \hline
Order & Variables & $R{^2}$ & r & $p-values$ & $F_{statistic}$ \\ 
  \hline
1 & FD2-IMF & 0.26 & -0.51 & 0.02 & 6.08 \\ 
  2 & FD2-SWS & 0.07 & -0.26* & 0.29 & 1.20 \\ 
  3 & FD2-Dst & 0.58 & 0.76 & $1.53\times 10^{-04}$ & 23.45 \\ 
  4 & FD2-ADV & 0.07 & 0.27* & 0.26 & 1.36 \\ 
   \hline
\end{tabular}
\end{table}

\begin{table}[ht]
\caption{ \textbf{ MTWS FD1 Regression Results with other Variables; $F_{critical}$ = 4.21; (95\% confidence). None statistically significant correlations are marked with *. Parameters are as defined in Table \ref{table 5}}}
\label{table 7}
\centering
\begin{tabular}{rlrrrr}
  \hline
 Order & Variables & $R{^2}$ & r & $p-values$ & $F_{statistic}$ \\ 
  \hline
1 & FD1-IMF & 0.10 & -0.32 & 0.09* & 3.03 \\ 
  2 & FD1-SWS & 0.19 & -0.44 & 0.02 & 6.44 \\ 
  3 & FD1-Dst & 0.55 & 0.74 & $4.75\times 10^{-06}$ & 32.42 \\ 
  4 & FD1-ADV & 0.13 & 0.36 & 0.06* & 3.99 \\ 
   \hline
\end{tabular}
\end{table}

\begin{table}[ht]
\caption{ \textbf{ MTWS FD2 Regression Results with other Variables; $F_{critical}$ = 4.38; (95\% confidence). None statistically significant correlations are marked with *. Parameters are as defined in Table \ref{table 5}}}
\label{table 8}
\centering
\begin{tabular}{rlrrrr}
  \hline
Order & Variables & $R{^2}$ & r & $p-values$ & $F_{statistic}$ \\ 
  \hline
1 & FD2-IMF & 0.02 & -0.14* & 0.55 & 0.36 \\ 
  2 & FD2-SWS & 0.17 & -0.41* & 0.07 & 3.81 \\ 
  3 & FD2-Dst & 0.44 & 0.66 & 0.001 & 14.98 \\ 
  4 & FD2-ADV & 0.0002 & 0.01* & 0.96 & 0.003 \\ 
   \hline
\end{tabular}
\end{table}

\onecolumn

\section*{Acknowledgements}
We feel indebted  to the group maintaining the website omniweb.gsfc.nasa.gov/html/ow data.html from where we downloaded the solar-geophysical data. The members of the R-mailing list (R-help@r-project.org) are gratefully acknowledged for assistance in codes for preliminary data analysis stage of the work. The invaluable contributions of the anonymous referee is hereby appreciated.


\bibliography{reference}

\section*{APPENDIX: Abbreviations and  Definitions}
\onecolumn

\begin{table}[ht]
\caption{Abbreviations and Definitions}
\label{Table A1}
\centering
\begin{tabular}{c c c }
  \hline\hline
S/N & Abbreviation & Definition\\ [0.5ex]
\hline
1 & CR & Cosmic ray \\
2 & FD & Forbush decrease\\
3 & ADV & Amplitude diurnal variation\\ 
4 & FD1 & Forbush decrease from filtered cosmic ray data\\ 
5 & FD2 & Forbush decrease from raw cosmic ray data\\
6 & IMF & Interplanetary magnetic field\\
7 & Dst & Geomagnetic storm index\\
8 & GCR & Galactic cosmic ray \\
9 & NMs & Neutron monitors\\
10 & IPDs & Interplanetary disturbances\\
11 & CMEs & Coronal mass ejections\\
12 & ICMEs & Interplanetary coronal mass ejections\\
13 & CHs & Coronal holes\\
14 & SEPs & Solar energetic particles\\
15 & GLEs & Ground level enhancements\\
16 & FFT & Fast Fourier transformation\\
17 & CLMX & Climax\\
18 & APTY & Apatity\\
19 & MTWS & Mt.Washington\\
20 & SWS & Solar wind speed\\
21 & MCMD & McMurdo\\
22 & THUL & Thule\\ [1ex]
\hline
\end{tabular}
\end{table} 

\end{document}